\documentstyle[12pt]{article}

\newcommand{\be}{\begin{equation}}
\newcommand{\ee}{\end{equation}}
\newcommand{\bea}{\begin{array}}
\newcommand{\ea}{\end{array}}
\newcommand{\beqa}{\begin{eqnarray}}
\newcommand{\eeqa}{\end{eqnarray}}
\newcommand{\bean}{\begin{eqnarray*}}
\newcommand{\eean}{\end{eqnarray*}}

\newcommand{\gapproxeq}{\lower .7ex\hbox{$\;\stackrel{\textstyle
>}{\sim}\;$}}
\newcommand{\lapproxeq}{\lower .7ex\hbox{$\;\stackrel{\textstyle
<}{\sim}\;$}}

\newcounter{appendice}

\def\thebibliography#1{{\bf REFERENCES\markboth
 {REFERENCES}{REFERENCES}}\list
 {[\arabic{enumi}]}{\settowidth\labelwidth{[#1]}\leftmargin\labelwidth
 \advance\leftmargin\labelsep
 \usecounter{enumi}}
 \def\newblock{\hskip .11em plus .33em minus -.07em}
 \sloppy
 \sfcode`\.=1000\relax}

\def\x{{\bf x}}
\def\z{{\bf z}}
\def\et{{\bf \eta}}
\def\BI{{\rm 1\!l}}
\def\Tone{\matrix{{}\cr T \cr {}^1}}
\def\Ttwo{\matrix{{}\cr T \cr {}^2}}

\def\up#1{\leavevmode \raise.16ex\hbox{#1}}

\def\sqr#1#2{{\vcenter{\vbox{\hrule height.#2pt
        \hbox{\vrule width.#2pt height#1pt \kern#1pt
          \vrule width.#2pt}
        \hrule height.#2pt}}}}

\def\BI{{\rm 1\!l}}

\begin{document}

\centerline{ \LARGE   Dirac Operator on the Quantum Sphere}

\vskip 2cm

\centerline{ {\sc   A. Pinzul and A. Stern }  }

\vskip 1cm
\begin{center}
 Department of Physics, University of Alabama,\\
Tuscaloosa, Alabama 35487, USA
\end{center}

\vskip 2cm

\vspace*{5mm}

\normalsize
\centerline{\bf ABSTRACT}

We  construct a Dirac operator on the quantum sphere $S^2_q$ which is covariant 
under the action of $SU_q(2)$.  It reduces to Watamuras' Dirac operator on
the fuzzy sphere when $q\rightarrow 1$. 
We argue that  our Dirac operator may  be useful
 in constructing $SU_q(2)$ invariant field theories
on $S^2_q$ following the Connes-Lott approach to noncommutative geometry.

\vskip 2cm
\vspace*{5mm}

\newpage
\scrollmode

Two important self-adjoint operators for the Connes-Lott approach
 to noncommutative geometry
are  the  
  Dirac operator and chirality operator.\cite{con},\cite{Mad},\cite{lan}
 The former is needed to construct a differential calculus and the latter
for chiral fermions.  If symmetries are present on
the   noncommutative manifold, then these operators should  respect the
symmetries if they  are to be
applied in writing invariant quantum field theories.
For example, the fuzzy sphere has an $SU(2)$ rotation symmetry which is reflected in the
Dirac operator.\cite{grpr},\cite{wat},\cite{BBIV},\cite{gkm}  
Other  noncommutative manifolds possessing
 different symmetry groups have been studied.\cite{bal2}
 On the other hand, the program of Connes has not been extensively applied to noncommutative
manifolds having symmetries  associated with quantum groups. 
(In this regard, see \cite{ps1}.)  In this letter we give a
construction of  the chirality operator  $\Gamma$ and Dirac operator $D$ 
for a noncommutative manifold having an $SU_q(2)$ symmetry.  The manifold is
Podles'   quantum sphere  $S^2_q$ .\cite{Pod}  It, along with the fuzzy sphere, has been
shown to appear in certain sectors of string theory.\cite{ars}   There are both
finite and infinite dimensional representations for the algebra depending
on the value of a certain parameter.  Finite 
 dimensional representations  were given in \cite{GMS}.
We shall give an explicit  construction of infinite dimensional representations  here.

 Besides the symmetry  requirements on $\Gamma$ and $D$, further conditions 
 are: a) that $\Gamma$ squares
to unity, b) $\Gamma$   commutes with the algebra 
${\cal A}$ associated with the noncommutative manifold, c)
  $\Gamma$ and $D$  anticommute and d) they yield the correct
 commutative limit.  To understand what the correct commutative limit is in our case,
 we shall first review the spherically symmetric  
Dirac operator on $S^2$.  We then generalize to  Dirac operators on a certain one parameter family of 
deformed (commutative) spheres.  These deformed spheres are one point compactifications of  
certain K\"ahler manifolds.  Their Dirac operators can also be regarded
as rotationally invariant, but now with respect to the Poisson action of $SU(2)$.
The Poisson action of $SU(2)$ on the deformed sphere
is the commutative limit of
 the action of the quantum group $SU_q(2)$ on  $S^2_q$.  So from our $D$ defined on 
$S^2_q$ we should recover the Dirac operator on the deformed sphere in the commutative limit.   
Although the latter Dirac operator  is rotationally invariant,
the property of invariance is  difficult to satisfy away from the commutative limit.
 Our $D$  on 
$S^2_q$  is instead   covariant.  We argue  that nevertheless 
it can be applied following
Connes' scheme for writing $SU_q(2)$ invariant field theories on $S^2_q$,
which is of current interest\cite{GMS}.  
Our construction of the
Dirac operator and chirality operator is along the lines of the Watamuras' construction
 for the fuzzy sphere\cite{wat}\footnote{Another Dirac operator on the
fuzzy sphere was given by Grosse and Pre\v{s}najder\cite{grpr}.},  and  in fact, their
 Dirac operator and chirality operator are obtained  from ours in a certain limit.
   \footnote{For a quite different construction, see 
\cite{OS}.  Also, a Dirac operator on $S^2_q$ was given in \cite{zum}
 which did not have simple $SU_q(2)$ transformation properties.  Its utility in
writing invariant theories is therefore unclear.}  From our $D$ one can thus construct
differential calculi over a family of noncommutative spheres (parametrized by $q$),
 including the fuzzy sphere.

\bigskip

${\bf S^2}\;.\quad $  The Dirac operator $\tilde D$ on $S^2$ can be expressed in different ways.\cite{jay}
 In terms of  stereographically projected coordinates 
$z$ and $\bar z = z^*$, it, along with the chirality operator $\tilde \gamma$, is given by  
 \be \tilde D  = \eta^{-3/2} \pmatrix{ & \partial \cr - \bar \partial & \cr}\eta^{1/2}
  \;,\qquad
 \tilde \gamma  = \pmatrix{1 &\cr & -1} \;, \label{tilddg}\ee
 where $\eta$ is the conformal factor
 $\eta = (1+|z|^2)^{-1}$ , $\partial= \frac{\partial}{\partial z},
\;\bar\partial= \frac{\partial}{\partial \bar z}$ and we assume unit radius.
$\tilde D$ and $ \tilde \gamma$  are hermitean and anticommute with each other.
Alternatively, the Dirac operator and chirality operator
 can be expressed in  a rotationally invariant manner
by  going to three-dimensional  embedded coordinates $x_i$,  $x_ix_i=1$,
 $i=1,2,3$ .  For this one can apply a unitary transformation  using 
\be U=\eta^{1/2} \pmatrix{ z & -1\cr 1 & \bar z \cr} \;,\label{untr}\ee  and
$z = (x_1 - ix_2)/ (2\eta)$ , $\bar z = (x_1 + ix_2)/ (2\eta)$ .  One gets
\beqa  D&=& U\tilde D U^\dagger= \sigma_i \ell_i +1\;,
\label{Dpos2} \\
 \gamma &=& U\tilde \gamma U^\dagger=  \sigma_i x_i    \;,\label{gons2}  \eeqa 
where $\ell_i$ are the angular momenta 
$ \ell_i =-i \epsilon_{ijk} x_j \frac{\partial}{\partial x_k}$  and $\sigma_i$ are Pauli matrices.
Yet another way to write $D$ is in terms of the rotationally
invariant  Poisson
structure on the sphere, i.e. \be \{x_i,x_j\}=\;\epsilon_{ijk} x_k  \;. \label{pbons2}\ee  
Acting  on a spinor $\psi$ it can be expressed by 
\be  -i \; D\psi  =  \{x\cdot \sigma, \psi\} 
 -\frac{\{x\cdot \sigma,\{x\cdot \sigma,x\cdot \sigma\}\}}{2\{
x\cdot \sigma,x\cdot \sigma\}} \;\psi \label{dopipb}\ee

\bigskip

${\bf Deformed }$   ${\bf S^2}\;.\quad $  We now generalize the notion of rotational invariance to the
include invariance with respect to Poisson actions of the rotation group.
In this approach $SU(2)$ is a Poisson-Lie group. 
The most general Poisson structure on $S^2$ where the left action of $SU(2)$  on $S^2$ 
is a Poisson action is given by\cite{SLW}
\be \{x_i,x_j\}=\;(1-\lambda
x_3)\; 
\epsilon_{ijk} x_k  \;, \label{lpbons2}\ee  where $ \lambda $ is a real constant
and we again assume $x_ix_i=1$. 
Substituting into (\ref{dopipb})\footnote{The inverse of $\{x\cdot \sigma,x\cdot \sigma\}$ is defined 
everywhere but $x_3= 1/\lambda$.}  gives the new Dirac operator
\be  D=(1-\lambda x_3)(\sigma_i \ell_i +1) + 
\frac {i\lambda}2 \epsilon_{ij3} x_i \sigma_j\;,
\label{lppb} \ee which is a one parameter deformation of (\ref{Dpos2}).
Like (\ref{Dpos2}), it is hermitean and anticommutes with the chirality operator (\ref{gons2}).  
It too is invariant under   simultaneous rotations of the spin and the coordinates,
 but now the latter is with a Poisson action.   If we perform a  unitary transformation
using the inverse of  (\ref{untr}) we recover (\ref{tilddg}) with the  conformal factor
 $\eta$  replaced by $\eta/(1- \lambda x_3)$ .   From (\ref{lpbons2})
 this factor also appears in the symplectic
two form, and as a result the manifold on which $D$ is written is K\"ahlerian, 
or more precisely
a one point compactification  of a K\"ahler manifold.  

\bigskip

${\bf S^2_q}\;.\quad $  For the  generalization to quantum groups, we recall that
Poisson actions are recovered from  actions of quantum groups in the commutative limit.
It is therefore natural to ask whether or not there is an extension of (\ref{lppb})
to a Dirac operator which is invariant under the action of a quantum group. 
As stated earlier, the invariance requirement appears difficult to satisfy.
On the other hand, we can find a Dirac operator with simple (covariant) transformation 
properties which is well suited for writing invariant field theories.
 The relevant
quantum group is $SU_q(2)$, because
 the Poisson action of $SU(2)$ is recovered from the action of $SU_q(2)$
when $q\rightarrow 1$.  $SU_q(2)$ has a natural action on the quantum sphere $S^2_q$ \cite{Pod},
which reduces to ordinary rotations on $S^2$  when $q\rightarrow 1$.  
We denote the generators of  
the algebra   ${\cal A}$  for 
$S^2_q$  by $\x_+$, $\x_-$  $\x_3$, along with the unit $1$. 
 They satisfy commutation
relations \beqa \x_+ \x_- - \x_- \x_+ &=& \mu \x_3 - (q-q^{-1})  \x_3^2 \cr
q\x_3 \x_+ - q^{-1} \x_+ \x_3 &=& \mu \x_+  \cr 
q\x_- \x_3 - q^{-1} \x_3 \x_- &=& \mu \x_- \;,\label{crfxs}\eeqa 
subject to the constraint
\be \x_3^2 +q \x_- \x_+ +q^{-1} \x_+ \x_- = 1 \label{radone} \ee
Not surprisingly, there are now two deformation parameters $q$ and $\mu$, which we take to
be  real (and we have again chosen the `radius' equal to one).  
 (\ref{crfxs}) and (\ref{radone}) are 
preserved under the involution $*$: $\x_{\pm}^* =\x_{\mp}$ and $\x_3^* =\x_3$.

There are two limits of interest of the relations  (\ref{crfxs}) and (\ref{radone}).
 The deformed sphere
is recovered when \be  q\rightarrow 1
 \;\; {\rm and} \; \; \mu \rightarrow 0\;,\qquad {\rm with}\;\; \frac{q-q^{-1}}{\mu}
 \rightarrow
 {\rm finite}\; ,  \label{clfqm} \ee
 which we refer to as the commutative limit.  
The limiting value of $(q-q^{-1})/{\mu}$ can be taken to be
 the constant $\lambda$ appearing 
in (\ref{lpbons2}) and (\ref{lppb}).   If we write $q=e^\tau$, then the 
 commutation relations (\ref{crfxs}) at lowest order in $\tau$ become 
\be [\x_i,\x_j] =    \frac{-2i\tau}{\lambda} \{\x_i,\x_j\}+ O(\tau^2) \;,\qquad
 i,j,...=1,2,3
\label{zocxx}\ee  where the Poisson brackets are those in (\ref{lppb}), and 
\be
 \x_1 = - \frac1{\sqrt{2}} (\x_+ + \x_-) \qquad  
\x_2 = - \frac{i}{\sqrt{2}} (\x_+ - \x_-)\;.  \label{x1x2}\ee
Here  $ \frac{-2\tau}{\lambda} $ plays the role of
$\hbar$. 
From (\ref{radone}) we get that $\x_i\x_i = 1+ O(\tau^2)$.

The other limit is \be q\rightarrow 1 \; \;{\rm and}\;\;
\mu \rightarrow  \pm\frac1{\sqrt{\ell(\ell+1)}}\;,\qquad \ell=\frac12, 1, \frac32, ...\;. 
\label{fzl}\ee
This is the limit of the fuzzy sphere associated with the $2\ell+1$ dimensional 
representation.  Now  (\ref{crfxs}) and (\ref{radone}) reduce to
the familiar relations\cite{Mad},\cite{grpr},\cite{wat},\cite{BBIV},\cite{gkm}  
\be [\x_i,\x_j] = -\frac i{\sqrt{\ell(\ell+1)}}\;
\epsilon_{ijk} \x_k\;,\qquad \x_i\x_i=1 \;, \ee
where we again define    $\x_1$ and $ \x_2 $ by (\ref{x1x2}).
\footnote{There exist deformations of these finite dimensional 
 representations when $q\ne 1$.  They occur for values of $\mu$ given in (\ref{dsfm}).
\cite{GMS},\cite{Pod}}
 If after taking the limit 
(\ref{fzl}) we then take $\ell \rightarrow \infty $ we get back  (undeformed) $S^2$.

From the   generators $\x_i$ it is convenient to defined  a $2\times 2$ 
 matrix $X=[X_{ab}]$ according to
\be X = \pmatrix{ q\x_3 & - \sqrt{ {\frac{[2]} q}} \x_+ \cr - \sqrt{{\frac{[2]}q}} \x_-
& -q^{-1} \x_3 \cr} \;, \ee    where $$[n]= {{1-q^{2n}}\over{1-q^2}}\;.$$  
$X$ has the properties: 

i) $X$ is hermitean with respect to
 $*$, i.e. $X_{ab}^*=X_{ba}$,

 ii) It satisfies a deformed trace condition:
$ {\rm Tr}_q X = q^{-1}X_{11} + q X_{22} = 0 \;, $

 iii)
$X^2 =\BI + \mu X \;,$ where  $\BI $ is the unit matrix.\footnote{Using this
property one can construct projection operators $P_\pm=\frac12 \{\BI \pm (X-\frac{\mu}2)/
\sqrt{1+\frac{\mu^2}4} \;\}$.  They are the  magnetic monopoles projectors of
\cite{TB}.  (We thank Tomasz Brzezinski for this remark.) They reduce to  
the magnetic monopoles projectors for the fuzzy sphere\cite{BBIV} in the limit
(\ref{fzl}).}

\noindent The matrices $X$  define an $SU_q(2)$ bimodule.
 $SU_q(2)$ matrices satisfy the   commutation
relations with themselves \be  R \Tone \Ttwo =\Ttwo \Tone R \label{RTT}
\ee   where
$\Tone = T \otimes \BI$,   $\Ttwo = \BI \otimes T$ and
\be  R=\pmatrix{q & & & \cr & 1 & & \cr & q-q^{-1} &1 & \cr
& & & q \cr}\;, \ee  and $R$ fulfills the quantum Yang-Baxter equation.
In addition, $T$ satisfies a unitarity condition (using the involution $*$)
 and also a deformed unimodularity condition
$det_q T=1$,  where $det_q T=T_{11}T_{22}
-qT_{12}T_{21}$.  This constraint
is possible because $det_q T$ so defined
is in the center of the algebra.  
Under the action of $SU_q(2)$ $X$ undergoes a
similarity transformation
\be X \rightarrow X'=TXT^{-1} \;,\label{alho}\ee which  preserves the
relations i)-iii).  Thus 
(\ref{alho}) is an algebra homomorphism.  Although matrix elements of $T$ do not
commute amongst themselves, they are assumed to commute 
with ${\cal A}$.  There is an analogue of the cyclic property of the trace 
(now with respect to 
$ {\rm Tr}_q $) for the matrices $T$ and this leads to  ii) being
 preserved under $SU_q(2)$.  

In either the commutative limit or the fuzzy limit,
 $X$ reduces to $\sigma_i x_i$, and the analogue of 
 transformation (\ref{alho}) rotates either the coordinates or the spin matrices:
\be \sigma_i x_i \rightarrow g\sigma_i g^\dagger \;x_i = \sigma_j \; \theta_{ji} x_i \;,
\ee  where $g\in SU(2)$ and $\theta$ is the corresponding rotation matrix.
Alternatively, we can write
\be \sigma_i x_i = g\sigma_i g^\dagger  \;\theta_{ij}^{-1} x_j \; \;,\label{invcond}\ee and
 say that $\sigma_i x_i$ is invariant with respect
to simultaneous rotations of the coordinates and the spin generated by the total
angular momentum.  There is no analogue of this statement with
regard to $SU_q(2)$ transformations of $X$.   For this define the deformed
Pauli matrices:
\be  \sigma^q_3 = \pmatrix{q &\cr &  q^{-1}\cr} \qquad 
\sigma^q_ + = \pmatrix{  & - \sqrt{ {\frac{[2]} q}}  \cr &  \cr} \qquad \sigma^q_ -= 
\pmatrix{  &  \cr - \sqrt{{\frac{[2]}q}} &  \cr} \;,\ee and write
$X= \sigma^q_3  \x_3 + \sigma^q_+\x_+ + \sigma^q_-   \x_- =\sigma^q_i  \x_i $.
Then (\ref{alho}) becomes \be \sigma^q_i  \x_i  \rightarrow T\sigma^q_i  T^{-1}\; \x_i =
 \sigma_j^q \; \Theta_{ji} \x_i \;,
\ee  
$ \Theta_{ij} $ giving the spin one representation of $SU_q(2)$ and $\x\rightarrow  
\Theta_{ij} \x_j $ defines the algebra homomorphism for $S^2_q$.  The analogue
of (\ref{invcond}) is true, namely,
\be \sigma_i^q \x_i = T\sigma_i^q T^{-1}  \;\Theta_{ij}^{-1} \x_j 
\; \;,\label{dinvcond}\ee  but $\Theta_{ij}^{-1}  $ is not in
  the spin one representation of $SU_q(2)$, and    $\x\rightarrow  
\Theta_{ij}^{-1} \x_j $ is not an algebra homomorphism for $S^2_q$.  
   For this reason
we cannot argue that $X$ is 
invariant under simultaneous $SU_q(2)$ transformations
 of the coordinates and the spin.  We can only
regard it as covariant with  respect to transformations of
 either the noncommuting  coordinates or the deformed spin matrices.  Since
 we will construct the
Dirac operator and chirality operator for $S^2_q$ from $X$,
the same will apply for these operators. 
  
In analogy with the Watamuras' construction
 for the fuzzy sphere, \cite{wat}   we introduce 
an ${\cal A}$-bimodule  ${\cal M}$, whose elements belong
to ${\cal A}$. 
  It is acted on from the left by
 ${\cal A}_L$ and the right by ${\cal A}_R$.  Elements   $a_L$, $b_L$,... of  
${\cal A}_L$ satisfy the same algebra as  ${\cal A}$, i.e. $a_L b_L= (ab)_L$,
$a,b,..\in {\cal A}$, while elements   $a_R$, $b_R$,... of  
${\cal A}_R$ satisfy $a_R b_R= (ba)_R$. 
 The action of an element $a_L\otimes b_R$ of
 ${\cal A}_L\otimes{\cal A}_R$ on   ${\cal M}$ 
is given by \be (a_L\otimes b_R) \;\circ\; {\cal M} = a {\cal M} b \ee It follows that left and
right variables commute, i.e. $[a_L\otimes 1\;,\;1\otimes b_R ]=0$.
Then we can fulfill the requirement that $\Gamma$ commutes with  ${\cal A}= {\cal A}_L$
by constructing it from only from elements of  ${\cal A}_R$ . 

Next we introduce a spin structure.  We define spinor fields $\Psi$ to  take
 values in ${\cal A}\otimes C^2$,
where $C^2$ is the space of two dimensional spinors, and to transform
covariantly under $SU_q(2)$, i.e.
\be \Psi \rightarrow \Psi ' = T \Psi \;,\label{traps}\ee 
They are acted on by operators ${\cal O}$  belonging to ${\cal A}_L\otimes{\cal A}_R\otimes M^2$,
where $M^2$ is the space of $2\times 2$ matrices.  For the result to be a spinor we need
that \be  ({\cal O}\circ \Psi) \rightarrow  {\cal O}'\circ\Psi ' =T ({\cal O}\circ \Psi) 
 \label{covreq}\ee
under $SU_q(2)$.  It is easy to find a solution to (\ref{covreq}) for arbitrary spinors
 for the case where ${\cal O}$ has only trivial dependence in ${\cal A}_R$.  Then
one can just write ${\cal O}={\tt L}$, where  \be {\tt L} \circ \Psi = X\Psi \label{lftac}\ee and use (\ref{alho}).
For the case where ${\cal O}$ has only trivial dependence in ${\cal A}_L$, 
one can  define ${\cal O}={\tt R}$, where
 \be ({\tt R} \circ \Psi)^T = \Psi^T\epsilon X \epsilon \;,\label{rftac}\ee 
The superscript $T$ denotes transpose and
\be \epsilon = \pmatrix{ & 1\cr -q &\cr} \;.\ee  Using  (\ref{alho}) and the identities
\be T\epsilon T^T = T^T \epsilon T = \epsilon \ee one can show that $X_R$ satisfies
the covariance condition (\ref{covreq}):
\beqa ({\tt R}'\circ \Psi')^T &=& \Psi'^T\epsilon X' \epsilon  \cr
&=& \Psi^T T^T \epsilon TXT^{-1} \epsilon  \cr
&=& \Psi^T  \epsilon X \epsilon T^T \cr
&=& (\;T\;({\tt R}\circ \Psi)\;)^T \eeqa 
From (\ref{lftac}) and (\ref{rftac}), ${\tt L}$ and  ${\tt R}$ have matrix elements
\be  {\tt L}_{ab} = (X_{ab})_L \qquad  {\tt R}_{ab} = \epsilon_{bc}\epsilon_{da}  
(X_{cd})_R   \ee
 Finally, one can construct $SU_q(2)$ covariant operators ${\cal O}$ with a nontrivial
 dependence in both ${\cal A}_L$  and ${\cal A}_R$ by taking matrix products
of ${\tt L}$ and ${\tt R}$.

We can now define the 
 chirality operator $\Gamma$.  $\Gamma$ is defined to square
 to the identity and commute with ${\cal M}$.  As remarked earlier, 
$\Gamma$ should be trivial in ${\cal A}_L$.  In addition,
we  require $\Gamma$ to satisfy the covariance condition  (\ref{covreq})
with respect to $SU_q(2)$ transformations. 
The solution for $\Gamma$  is then
\be \Gamma =\frac1{  q\sqrt{4+\mu^2}   }\;(2{\tt R}+{q\mu}  \BI)  \;. \ee 
Its matrix elements are
\be \Gamma_{ab} = \frac1{ q\sqrt{4+\mu^2}   }\;(2 \epsilon_{bc}\epsilon_{da}  
(X_{cd})_R  +{q\mu} \delta_{ab})  \;. 
\label{gme}\ee
 
The Dirac operator $D$ is required to anticommute with $\Gamma$.  We shall also
demand that it 
 transform covariantly under $SU_q(2)$.
 These  requirements are met
for any $D$ of the form 
 \be D =\Gamma\;[Y,\Gamma]\;, \ee where  $Y$ 
 transform covariantly 
under $SU_q(2)$.  For $D$ to have a nontrivial commutator with ${\cal M}$, $Y$
should be nontrivial in ${\cal A}_L$.
A natural choice is therefore 
\be  Y={1\over {2\mu}} \;   {\tt L}\;,\ee  The factor $1\over {2\mu }$ was inserted
to get a finite commutative limit, defined in (\ref{clfqm}), which we compute below.
We show that if we once again choose
the limiting value of $(q-q^{-1})/{\mu}$ to be the constant $\lambda$ appearing 
in (\ref{lpbons2}) and (\ref{lppb}) we recover the Dirac operator (\ref{lppb})
 for the deformed
sphere.  If on the other hand we take the fuzzy sphere limit
(\ref{fzl}), $D$ reduces to the Dirac operator in \cite{wat}.
For arbitrary values of $q$ and $\mu$, the  matrix elements of $D$ are
\be D_{ab} ={1\over {2q^2\mu (4+\mu^2)}} \;  
 (X_{cd})_L\;\otimes \;\biggl(
(2\epsilon X\epsilon +{q\mu}  \BI)_{bd} \;(2\epsilon X\epsilon +{q\mu}  \BI)_{ca}-
 q^2(4+\mu^2)\delta_{bd} \delta_{ca} \biggr)_R  \;. \ee

We now show that the  commutative limit of $\Gamma$ and $D$ is (\ref{gons2}) 
and (\ref{lppb}), respectively.  As earlier, we set $q=e^\tau$ and $\mu=2\tau/\lambda 
+O(\tau^2)$
and then perform an expansion in $\tau$.  Up to first order
\beqa  X &=& \x_i  \sigma_i + \tau \x_3 \BI + O(\tau^2) \cr & & \cr
\epsilon X \epsilon &=& \x_i  \sigma_i^T  +\tau \; \biggl( \x_3(\sigma_3- \BI)
 + (\x_1-i\x_2)(\sigma_1 -i\sigma_2)\biggr) + O(\tau^2) \;, \eeqa
where $\sigma_i$, $i=1,2,3$, are Pauli matrices and  $\x_1$ and $ \x_2 $ were
defined in (\ref{x1x2}).  Then
\be\frac1{ q\sqrt{4+\mu^2}   }\;(2\epsilon X\epsilon +{q\mu}  \BI) = \x\cdot \sigma^T 
+ \tau \;\biggl(i\epsilon_{ij3} \x_i \sigma_j^T
+ (\lambda^{-1}-\x_3)\BI \biggr)+ O(\tau^2) \;. \ee 
We only need the zeroth order term 
to show that the commutative limit of $\Gamma$ is (\ref{gons2}),   
while we need the first order term to compute $D$. 
 Up to first order in $\tau$, $D$ acting on a spinor $\psi$ is given by
\beqa 2\mu \;D_{ab} \circ \psi_b &\rightarrow  &
 (\x\cdot \sigma)_{cd} \;\psi_b \;(\x\cdot \sigma)_{db} \;(\x\cdot \sigma)_{ac}
- (\x\cdot \sigma )_{ab}\;\psi_b
\cr & & \cr & & +\;2\tau \biggl( i  \epsilon_{ij3} \x_i \sigma_j \psi +                  
  (\lambda^{-1}-\x_3) \;\psi\biggr)_a \;,\label{clod}  \eeqa  as $q\rightarrow 1$. 
 We have neglected the ordering of factors of $\x_i$ and $\psi_a$ in the first order terms,
which is not valid at zeroth order.  For the latter, we can use (\ref{zocxx}) with
 Poisson brackets given in (\ref{lppb}).  This gives
\be 2\mu \;D_{ab} \circ \psi_b \rightarrow  4\tau ( \lambda^{-1}-\x_3)
(-i \epsilon_{ijk} \sigma_i\x_j \partial_k \psi + \psi )_a  
+2\tau i \epsilon_{ij3} \x_i(\sigma_j \psi )_a 
 \;, \ee and  consequently (\ref{lppb}).  

In order to proceed with  Connes' construction of the differential calculus,
one must obtain the spectra
of the Dirac operator and introduce a  Hilbert space for the spinors $\Psi$. 
 We have not yet attempted the former.  Concerning the latter, 
there are both finite and infinite dimensional Hilbert spaces.
The  finite  dimensional Hilbert spaces occur for
certain discrete values of $\mu$:
\be \mu= \pm \frac{[2(2\ell+1)]}{q\;[2\ell+1]}\; \frac1{\sqrt{[2\ell][2\ell+2]}}\;,
 \qquad \ell=\frac12, 1, \frac32, ...\;, \label{dsfm} \ee  where the dimension is
$2(2\ell +1)$.  (The factor $2$ is because we have spinors.)
They were explicitly constructed in \cite{GMS}.
In the limit  $q\rightarrow 1 $, (\ref{dsfm}) goes to (\ref{fzl}) and
 the corresponding  matrix
 representations for the algebra agree with those of the fuzzy sphere.

Infinite dimensional representations occur for
\be \mu = q- q^{-1} \;,\label{svfm}\ee  which can emerge as the large $\ell$
limit of (\ref{dsfm}).    
A possible construction involves the use of coherent
states.  Here one can adopt the approach developed in \cite{mmsz},\cite{us} which
relies on coherent states for deformed creation and annihilation operators, 
 $\tilde {\bf a}^\dagger$ and $\tilde {\bf a}$, respectively.  They  satisfy
the commutation relations  \be [\tilde{\bf a},\tilde{\bf a}^\dagger]=\;
F(\tilde {\bf a}\tilde
{\bf a}^\dagger)
\label{capaad}\ee for some function $F$.  The deformed coherent states diagonalize
$\tilde {\bf a}$, and  have a natural scalar product. 
 An explicit construction was given in \cite{us} 
 for the fuzzy sphere.  There one identified 
$\tilde {\bf a}^\dagger$ and $\tilde {\bf a}$ with the fuzzy analogues of
stereographic coordinates.  Something similar can be done for the quantum sphere,
since  analogues of
stereographic coordinates also exist for $S^2_q$. \cite{zum}  This  construction,
however, only works when (\ref{svfm}) is satisfied.
 Then one can parametrize the matrix $X$ according to 
\be X = \pmatrix {q(1-[2]\et) & q^{-1}[2] \z \et\cr 
q^{-1}[2]  \et\bar \z & q^{-1}([2] \et - 1)} \;,\label{pfx}\ee where $\bar \z = \z^*$, and we assume that the operator $1+\bar \z \z $
is nonsingular, with \be \et^{-1} = 1+\bar \z \z = q^2( 1+\z\bar  \z)\;,\ee   which
gives the commutation relations for $\z$ and $\bar \z$.  From these relations
 one can verify i-iii).  Now  identify:  $\tilde {\bf a}=\z$ , $\tilde {\bf a}^\dagger
=\bar\z$ and $F(\z\bar \z) = -q^{-1} \mu \et^{-1} $, where we used (\ref{svfm}).
To write the deformed coherent states one introduces a  map from a pair of 
standard (or undeformed) creation and annihilation operators, 
 $ {\bf a}^\dagger$ and $ {\bf a}$,  satisfying
  \be [{\bf a},{\bf a}^\dagger]=\;1\label{udcapaad}\;,\ee to 
$ \tilde{\bf a}^\dagger$ and $ \tilde {\bf a}$.  
The map can be  expressed as
  \be \tilde {\bf a}= f({\bf  n}+1)\; {\bf a }\;,\label{atfa}\ee
  ${\bf  n}$ being
 the number operator ${\bf  n}={\bf a}^\dagger {\bf a}$,
having  eigenvalues $n=0,1,2,...$ .  The function $f$ is
determined from $F$.    We get
\be |f(n)|^2 = - q^{-2n+1} \mu  \frac{[n]}{n}\;,n>0\;.\label{fnsq}\ee 
For $n=0$ we can take the limiting value $|f(0)|^2 = -2\ln q $ .
For the right hand side of (\ref{fnsq}) to be positive we must restrict $q<1 $.
We note that the undeformed  creation and annihilation operators
are not recovered in the commutative limit, i.e. $q\rightarrow 1$,
and further $f$ is ill-defined in the limit.
The Hilbert space ${\tt H}$ can now be defined as being
spanned by the eigenstates 
$|n> $, $n=0,1,2,...$, of ${\bf n}$ with
scalar product $<n|m>=\delta_{n,m}$.  (Actually, we want two copies of ${\tt H}$ 
for the spinors.)
Alternatively, one can use the overcomplete coherent state basis: \cite{mmsz},\cite{us} 
\beqa |\zeta > &=& 
 N (|\zeta |^2  )^{-\frac12}\; \exp{\{\zeta f({\bf n})^{-1}{\bf a}^\dagger\}}\;
f({\bf n})^{-1} \;|0> \cr  & &\cr  &=& 
  N (|\zeta |^2  )^{-\frac12}\;\sum_{n=0}^{\infty} \frac{\zeta^n }{\sqrt{n!}\;
[f(n)]!} \;|n> \;,\label{ndcs}\eeqa
where  $[f(n)]!= f(n)f(n-1)...f(0)$, which  diagonalize $\tilde {\bf a}$.
 Requiring  $ |\zeta > $ to be of unit norm fixes $N (|\zeta |^2  )  $,
\be N (x  )  =  \sum_{n=0}^{\infty} \frac{x^{n} }{n!\;([f(n)]!)^2} \;. \ee
As with the standard coherent states, the states (\ref{ndcs}) are  not orthonormal,
but instead satisfy
\be <\eta | \zeta >  =   N (|\eta |^2  )^{-\frac12} \; N (|\zeta |^2  )^{-\frac12}
 N(\bar\eta\zeta)   \label{zetet} \ee
(Alternatively, a construction of finite dimensional Hilbert spaces exists for 
certain discrete values of $\mu$ \cite{GMS}.) 

 From the above, we conclude that when  (\ref{svfm}), along with $q<1$, are satisfied
there is an infinite dimensional Hilbert space which should allow for a
construction of a differential calculus on $S^2_q$ following Connes.
We recall that  the condition (\ref{svfm}) 
was previously found to be necessary for writing down  a  differential calculus on $S^2_q$      
in a very different approach.  \cite{pod2}   
The  differential calculus of  \cite{pod2} was obtained by demanding invariance of
the exterior derivative $d$
under $SU_q(2)$.  The corresponding algebra of one forms is easily  expressed 
 in terms of $\z$ and $\bar \z$ and their exterior derivatives:\cite{zum}
$$ \z d\z =q^{-2} d\z \z \;,\qquad \z d\bar \z =q^{-2} d\bar \z \z   $$
$$ \bar \z d\z =q^{2} d\z \bar \z \;,\qquad \bar \z d\bar \z =q^{2} d\bar \z \bar \z   $$
\be d\z d\bar \z =-q^{-2} d\bar \z d\z \;,\qquad (d\z )^2  = (d\bar \z )^2  = 0 \label{pdc}
\ee
We, on the other hand, do not recover these formulae upon representing 
$d\z $ and $d\bar \z$ with $ [D,\z]$ and $ [D,\bar \z]$, respectively, following Connes.
 This cannot be  surprising since  our
Dirac operator  is only covariant and  therefore, in contrast to \cite{pod2}, will
 not lead to an invariant exterior
derivative.  (From the Dirac operator of \cite{zum} one does recover the Podles
differential calculus (\ref{pdc}), however, that Dirac operator does  not have simple
transformation properties under $SU_q(2)$.)

Although our exterior derivative is not invariant, it should  nevertheless be useful
for writing invariant field theories\cite{GMS}.  
For example, if $\phi\in {\cal A}$ represents
a scalar field on $S^2_q$, we can construct the following  quadratic invariant
\be {\rm Tr}_q [D,\phi]  [D,\phi]  \;, \ee
since ${\rm Tr}_q T{\tt M}T^{-1}={\rm Tr}_q {\tt M }$ for  any ${\tt M}
\in {\cal A}\times M^2$. 
 For fields belonging to nontrivial representations of $SU_q(2)$ additional
deformed traces can be taken.  In order to  be
useful in constructing actions, we will also have to take a trace in the 
 Hilbert space, which will require clarification.
  More exciting is the possibility that $q$ can be made into
a dynamical quantity, thereby introducing dynamics in the underlying 
noncommuting manifold, and possibly allowing for quantum fluctuations about fuzzy spheres.

\bigskip
\noindent
{\bf Acknowledgement}

\noindent
This work was supported by the joint NSF-CONACyT grant
 E120.0462/2000, and 
 in part by the U.S. Department of Energy
 under contract number DE-FG05-84ER40141.

\end{document}